\documentclass[12pt]{article} 
\usepackage{amsmath, amssymb, graphicx, url, inputenc, fontenc, hyperref, amsfonts, bm, bbm, mathdots, arydshln, multirow, multicol, algorithm, algorithmic, setspace, authblk}
\usepackage[margin=1in]{geometry}
\usepackage[square, numbers]{natbib}

\title{Sparse and Low-bias Estimation of High Dimensional Vector Autoregressive Models}
\date{}

\author[1]{Trevor D. Ruiz\thanks{Corresponding Author: trevor.ruiz@oregonstate.edu}}
\author[1]{Sharmodeep Bhattacharyya}
    \affil[1]{Department of Statistics, Oregon State University, Corvallis, OR}

\author[2]{Mahesh Balasubramanian}
    \affil[2]{Compiler Microarchitecture Lab, Arizona State University, Tempe, AZ}

\author[3]{Kristofer Bouchard} 
    \affil[3]{Biological Systems and Engineering Division, Lawrence Berkeley National Labs, Berkeley, CA} 

\begin{document}

\maketitle\bigskip

\begin{abstract}%
Vector autoregressive ($VAR$) models are widely used for causal discovery and forecasting in multivariate time series analysis. In the high-dimensional setting, which is increasingly common in fields such as neuroscience and econometrics, model parameters are inferred by $L_1$-regularized maximum likelihood (RML). A well-known feature of RML inference is that in general the technique produces a trade-off between sparsity and bias that depends on the choice of the regularization hyperparameter. In the context of multivariate time series analysis, sparse estimates are favorable for causal discovery and low-bias estimates are favorable for forecasting. However, owing to a paucity of research on hyperparameter selection methods, practitioners must rely on \textit{ad-hoc} methods such as cross-validation (or manual tuning). The particular balance that such approaches achieve between the two goals --- causal discovery and forecasting --- is poorly understood. Our paper investigates this behavior and proposes a method ($UoI_{VAR}$) that achieves a better balance between sparsity and bias when the underlying causal influences are in fact sparse. We demonstrate through simulation that RML with a hyperparameter selected by cross-validation tends to overfit, producing relatively dense estimates. We further demonstrate that $UoI_{VAR}$ much more effectively approximates the correct sparsity pattern with only a minor compromise in model fit, particularly so for larger data dimensions, and that the estimates produced by $UoI_{VAR}$ exhibit less bias. We conclude that our method achieves improved performance especially well-suited to applications involving simultaneous causal discovery and forecasting in high-dimensional settings.
\end{abstract}

\noindent 
{\it Keywords:} multivariate time series, vector autoregressive models, high dimensional data, sparsity, LASSO, Granger causality, graphical models, bootstrap, Union of Intersections%
\vfill

\newpage
\spacing{1.5}
\section{Introduction}

Multivariate time series data are now ubiquitous across scientific fields and increasingly high-dimensional. In neuroscience, for instance, intra-cortical electrophysiology produces simultaneous recordings of neural activity as measured by large arrays of hundreds to thousands of electrodes at sampling rates exceeding 24kHz \citep{buzsaki2012}. Data with analogous structure is generated in neuroscience from electroencephalography (EEG) \citep{astolfi2007comparison}, and various other sources \citep{BKM04, pillow2008, bassett2011dynamic}. In econometrics and fincance, multivariate time series data is used for forecasting, macroeconomic studies, and structural analysis \citep{sims1980macroeconomics, fackler1986application, forni2005generalized, stock2002forecasting, tsay2005analysis}. Similar data are arising on increasing scales in environmental science and geosciences \citep{karpatne2018, triacca2013}, epidemiology \citep{cunniffe2015}, and sociology \citep{mcfarland2016}. Such growing data dimensionalities and length of recordings present opportunities for scientific discovery alongside methodological and computational challenges for data analysis.


Vector autoregressive ($VAR$) models provide a flexible framework for forecasting, structural analysis (finding a unique process parametrization under constraints on the error term), impulse response analysis (describing the propagation of a `shock' or erratic event throughout the system), and estimation of various types of causality \citep{lutkepohl2005}. Furthermore, $VAR$ model parameters are conceptually straightforward to estimate, although computationally scaling to large systems remains a challenge.


In particular, large datasets require high-dimensional process models. In this context, parameters are estimated with sparsity-inducing constraints, which has motivated research on sparse estimation of high-dimensional vector autoregressive model parameters \citep{song2011large, fan2011sparse, han2015direct, qiu2016joint, basu2015, hall2016}. Interesting sparsity constraints also arise in related literature on joint estimation of multiple Gaussian graphical models \citep{guo2011joint, danaher2014joint}.


Sparse estimation methods for time series models typically rely on $L_1$-regularized maximum likelihood estimation. However, it is known that this technique can result in overfitting --- specifically overly-dense estimates --- and excessive bias \citep{buhlmann2011, meinshausen2010} in high-dimensional regression and precision matrix estimation, and it is likely that these problems persist in the time series context. Nonetheless, few alternatives to RML are available to date in multivariate time series analysis for high-dimensional data. 


This paper offers a two-fold contribution to the above-cited work: (i) we propose an inference procedure ($UoI_{VAR}$) that improves on $L_1$-regularized estimation (RML) of high-dimensional vector autoregressive models by leveraging the Union of Intersections ($UoI$) statistical machine learning framework \citep{bouchard2017}; and (ii) we provide simulation-based and theoretical support for the algorithm. Additionally, we compare $UoI_{VAR}$ to RML in an example Granger causality analysis identifying mutual influences among the share prices of companies on the S\&P 500 index. Together, these results indicate that $UoI_{VAR}$ will enhance inference in $VAR$ models across application domains.

\section{Methods}

\subsection{Vector autoregressive models}
The stochastic process $\{X_t : \Omega \rightarrow \mathbb{R}^M\}_{t \in \mathbb{Z}}$ is a vector autoregressive process of order $D$ ($VAR(D)$) if for all $t \in \mathbb{Z}$
\begin{equation} 
X_t = \nu + \sum_{d = 1}^D A_d X_{t - d} + \epsilon_t
\;,\quad 
\begin{cases}
\mathbb{E}\epsilon_t &= 0 \\
\mathbb{E}\epsilon_t\epsilon_t' &= \Sigma \\
\mathbb{E}\epsilon_t\epsilon_s' &= 0
\end{cases}
\label{eqn:vard}
\end{equation}
where $\Sigma \in \mathbb{R}^{M \times M}$ is positive definite and $A_1, \dots, A_D$ satisfy $\text{det}\left( I - \sum_{d = 1}^D A_d z^d \right) \neq 0$ for all $|z| \leq 1$. The latter condition ensures that the process is well-defined, stationary and stable.

Given an observed time series of length $T$, denoted $\{ x_t \in \mathbb{R}^M \}_{t = 0}^T$, a $VAR(D)$ model for the data is simply some $VAR(D)$ process stipulated as the data-generating mechanism. The model can be expressed in the form of a multivariate multiple regression $Y = UB + E$ where the response is denoted by $Y = (x_T \; x_{T - 1} \; \cdots \; x_D)^T$ comprises the observations beginning from time $D$, the error terms are denoted by $E = (\epsilon_T \; \epsilon_{T - 1} \; \cdots \; \epsilon_D)^T$, and the linear predictor $\bm{UB}$ on the right-hand side denotes 
\begin{equation*}
\label{eqn:var2}
\left(\begin{array}{c:c:c:c}
	1 &x_{T - 1}^T &\cdots &x_{T - D}^T \\
	1 &x_{T - 2}^T &\cdots &x_{T - D - 1}^T \\
	\vdots &\vdots &\ddots &\vdots \\
	1 &x_{D - 1}^T &\cdots &x_{0}^T \\
	\end{array}\right)
	\left(\begin{array}{c}
	\nu^T \\
	A_1^T \\
	\vdots \\
	A_D^T
	\end{array}\right)
\end{equation*}
The classical estimation technique is to estimate $B$ with ordinary least squares (OLS) by $\hat{B} = (U'U)^{-1}U'Y$, and then estimate $\Sigma$ by $\hat{\Sigma} = \frac{1}{T - 1}(Y - U\hat{B})(Y - U\hat{B})'$. The equivalence of this procedure with maximum likelihood estimation is well-established \citep{lutkepohl2005}.

When $M$ is large and $A_1, \dots, A_D$ are sparse, $B$ is instead estimated using $L_1$-regularized maximum likelihood (RML) by
\begin{equation}
\hat{B} = \arg\min_{B} \left\{ 
			-\ell(B;X) + \lambda P(B)
			\right\}
\label{eqn:var_lasso}
\end{equation}
where $\ell(B;X)$ denotes the log-likelihood of $B$ given data $X = (x_1 \cdots x_T) \in \mathbb{R}^{M \times T}$, and $P(B) = \|\text{vec}B\|_1$ is the sparsity-inducing regularization term.\footnote{Due to the equivalence between maximum likelihood and ordinary least squares, $\|Y - UB\|_F^2$ can be substituted for $-\ell(B; X)$ in Equation (\ref{eqn:var_lasso}) and LASSO regression on the vectorized problem (obtained by column-stacking the response $Y$) can be used to find the solution with fast, numerically stable, and widely available algorithms \citep{friedman2007, friedman2010}.}

\subsection{Union of Intersections algorithmic framework}

The Union of Intersections ($UoI$) statistical-machine learning framework that separates sparse parameter selection (an `intersection step') from parameter estimation (a `union' step). The advantages of this approach are established for several sparse learning techniques, including regression, classification, and matrix decomposition \citep{bouchard2017, ubaru2017}. The intersection step first infers several candidate parameter support sets (sets of nonzero parameter locations). This is accomplished through inferring support sets for a range of sparsity-inducing regularization strengths on bootstrap samples, and applying intersection operations across the bootstrap samples. In the union step, estimates are calculated without regularization for each candidate support on bootstrap samples, and predictive quality is evaluated on separate samples. Finally, the estimates that optimize predictive quality are averaged (a union operation with respect to the selected candidate supports) to produce the final output.

\subsection{Estimation algorithm}

Our estimation algorithm separates the estimation procedure into an intersection step utilizing RML and a union step utilizing OLS. The intersection step estimates a collection of candidate support sets for the transition matrices $A_1, \dots, A_D$ for a $VAR(D)$ model, and the union step produces the final estimates $\hat{A}_1, \dots, \hat{A}_D$. These steps are given in detail separately as Algorithms \ref{alg:intersection} and \ref{alg:union}.

In the intersection step (Algorithm \ref{alg:intersection}), supports of RML estimates are computed for a fixed regularization path $\lambda$ on $B_1$ bootstrap samples. These support sets are aggregated across bootstrap samples separately for each $\lambda_k$ by a thresholded intersection operation. In other words, this procedure selects consistently recurring RML support sets under resampling of the data $B_1$ times, with a specifiable recurrence threshold $s$.

\begin{algorithm}
\caption{Intersection step}
\label{alg:intersection}
\begin{algorithmic}
\REQUIRE{
	\STATE data $\{x_t \in \mathbb{R}^M\}_{t = 0}^T$ 
	\STATE regularization path $\lambda \in \mathbb{R}^K$ 
	\STATE number of bootstrap samples $B_1$
	\STATE thresholding parameter $s$
	}
\FOR{$b = 1$ to $B_1$}
\STATE draw bootstrap sample $\{x_t^*\}_{t = 0}^T$\;
\FOR{$k = 1$ to $K$}
\STATE $\tilde{B}_k \leftarrow \arg\min_{B}\left\{-\ell(B\;;\;x_0^*, \dots x_T^*) + \lambda_k P (B)\right\}$
\STATE $S_{b, k} \leftarrow \{(i, j):\; \tilde{b}_{ij} \neq 0\}$
\ENDFOR
\ENDFOR
\STATE $S_k \leftarrow \left\{ (i, j):\; \sum_{b = 1}^{B_1} \mathbbm{1}\{ (i, j) \in S_{b, k}\} \geq s B_1\right\}$
\ENSURE{Support sets $S_1, \dots, S_k$}
\end{algorithmic}
\end{algorithm}

The union step (Algorithm \ref{alg:union}) begins by repeating the following procedure $B_2$ times. Training and test bootstrap samples are drawn, and OLS estimates are computed for each candidate support set from the intersection step on the training bootstrap sample. The estimates that minimize a user-defined loss function $f$ on the test set are stored. Each iteration of the procedure is repeated with new boostrap samples, producing $B_2$ parameter estimates. Finally, an average over the $100\times q$\% most sparse estimates produces the output. This portion of the algorithm averages OLS estimates selected by bootstrapped cross-validation under resampling of the data $B_2$ times.

\begin{algorithm}
\caption{Union step}
\label{alg:union}
\begin{algorithmic}
\REQUIRE{
\STATE data $\{x_t \in \mathbb{R}^M\}_{t = 0}^T$
\STATE candidate support sets $S_1, \dots, S_K$ (from intersection step)
\STATE number of bootstrap samples $B_2$
\STATE loss function $f$
\STATE threshold parameter $q$
}
\FOR{$b = 1$ to $B_2$}
\STATE draw bootstrap samples $\{x_t^{*(1)}\}_{t = 0}^T$, $\{x_t^{*(2)}\}_{t = 0}^T$
\FOR{$k = 1$ to $K$}
\STATE $\mathcal{B} \leftarrow \left\{B \in \mathbb{R}^{(D + 1)\times M}: \mathbbm{1}\{b_{ij} \neq 0\} = \mathbbm{1}\{(i, j) \in S_k\}\right\}$
\STATE $\tilde{B}_k \leftarrow \arg\min_{B\in\mathcal{B}} \left\{-\ell(B\;;\;x_0^{*(1)}, \dots, x_T^{*(1)})\right\}$
\STATE $f_k \leftarrow f(\tilde{B}_k; x_0^{*(2)}, \dots, x_T^{*(2)})$
\ENDFOR
\STATE $\hat{B}^{(b)} \leftarrow \tilde{B}_J: J = \arg\min_{k} \{f_k\}$
\STATE $q_b \leftarrow \sum_{i, j}\mathbbm{1}\{\hat{b}_{ij}^{(b)} \neq 0\}$
\ENDFOR
\STATE $Q \leftarrow \{b: \sum_{j = 1}^{B_2} \mathbbm{1}\{q_j \leq q_b\} \leq q B_2\}$
\STATE $\hat{B} \leftarrow \frac{1}{|Q|} \sum_{b \in Q} \hat{B}^{(b)}$\;
\ENSURE{estimate $\hat{B}$}
\end{algorithmic}
\end{algorithm}

Given the intensive use of resampling methods, bootstrap procedures suitable for time series are required to implement this method. One of the most common bootstrap procedures for time series is the moving block bootstrap \citep{kunsch1989jackknife, liu1992moving}, in which the original time series is divided into overlapping blocks and the blocks are randomly sampled with replacement to construct a resampling of the original time series. The choice of appropriate block lengths is dependent on the statistical problem \citep{buhlmann1999block, kreiss2012}.

The hyperparameters $B_1$, $B_2$, $s$, and $q$ control the sparsity of the estimates at each stage of the algorithm. Increasing $B_1$ produces greater sparsity among candidate supports, decreasing $s$ creates less sparsity, and doing both simultaneously helps to stabilize the candidate supports against erratic bootstrap samples. Similarly, increasing $B_2$ generally increases density and decreasing $q$ counteracts this effect, and doing both stabilizes the output of the union step. The ability to tune the algorithm using $s, q$ can be especially helpful when computational resources are limited.

\section{Simulation study}

We assessed the performance of $UoI_{VAR}$ on synthetic data relative to RML for a range of data dimensionalities. Specifically, we conducted a simulation study utilizing synthetic datasets conforming to each combination of process dimensions $M = 5, 10, 20, 40, 50$ and time series lengths $T = 50, 100, 200$. The study examined selection accuracy, model fit, and estimation bias for each method.

For each simulation setting (combination of $M, T$), $VAR(1)$ process parameters were generated as follows: $M$ nonzero transition matrix parameters were drawn from a frequency distribution increasing exponenially away from zero in either direction; $\nu = 0$; and $\Sigma = 0.5I$. Thus, the parameters exhibit $1 - 1/M$\% sparsity. Nonzero parameter positions were randomly allocated to transition matrix positions. Then 50 realizations of each process were simulated, and model estimation was conducted using RML and $UoI_{VAR}$. Both methods utilized the same regularization path $\lambda$. For RML, the regularization strength that minimized the average of one-step and four-step forecasting errors over five-fold cross-validation was used for the final estimate. We used the same criterion for the loss funciton $f$ in our algorithm, along with hyperparameters $B_1 = 10$, $B_2 = 50$, $s = 1$, and $q = 0.3$.

\begin{figure}
    \centering
    \includegraphics[width = 1\linewidth]{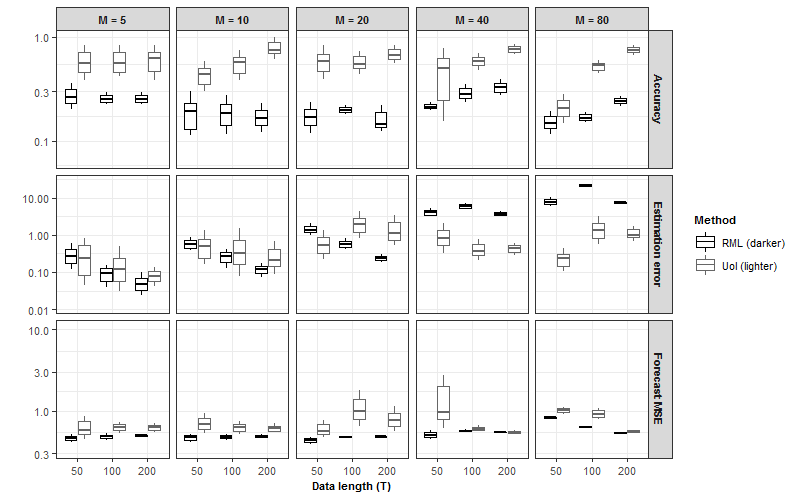}
    \caption{Selection, estimation, and forecasting behavior observed in the simulation study. Each row shows boxplots of a different metric computed from the estimates of each method for each of the 50 repetitions of the simulation setting; the rows are faceted according to the process dimension $M$, and each facet displays time series length $T$ on the horizontal axis. Top row, selection accuracy $1 - \frac{FN + FP}{M + FP}$; middle row, estimation error computed on the estimated support $\sum_{i, j \in \hat{S}}\|a_{ij} - \hat{a}_{ij}\|_2^2$; bottom row, one-step forecast error $\frac{1}{T - 1}\sum_t\|X_{t + 1} - \hat{X}_{t + 1}\|_2^2$.}
    \label{fig1}
\end{figure}

Figure \ref{fig1} summarizes the results of the simulation study. Selection accuracy is articulated as a combination of false positives $FP$, the number of positions in the parameter estimates that are in fact zero but estimated as nonzero, and false negatives $FN$, the number of positions that are in fact nonzero but estimated as zero; the accuracy metric is then $1 - \frac{FN + FP}{M + FP}$. Across all settings, RML exhibits low accuracy and $UoI_{VAR}$ exhibits improved accuracy; these behaviors are driven predominantly by false positive rates (not depicted in the figure). The main limitation of $UoI_{VAR}$ is an increased false negative rate relative to RML when less data are available (shorter $T$ settings). However, this problem diminishes rapidly as $T$ increases, and as a result, for larger $M$, the selection performance of our method improves much faster than RML as $T$ increases. Furthermore, as depicted in the second row of the figure, $UoI_{VAR}$ achieves dramatically lower estimation errors in large-$M$ settings. Finally, it appears that these improvements come at the cost of a slight decrease in fit; the third row shows forecasting errors on the data to which the models were fit.

\section{Data analysis}

To illustrate an application and further compare each method in the context of a data analysis, a $VAR(1)$ model was used to identify putative causal connections between weekly closes of 50 randomly chosen publicly traded companies listed on the S\&P 500 index in 2013-2014. This dataset was chosen due to the absence of benchmark datasets with known ground truth for large multivariate time series; the years 2013-2014 saw a steadily climbing index with no major disturbances. To obtain an approximately stationary process, first-order differences were calculated from the raw series; then, $VAR(1)$ model parameters were estimated from these differences using our method and RML.

\begin{figure}
  \centering
  \includegraphics[width = 0.9\linewidth]{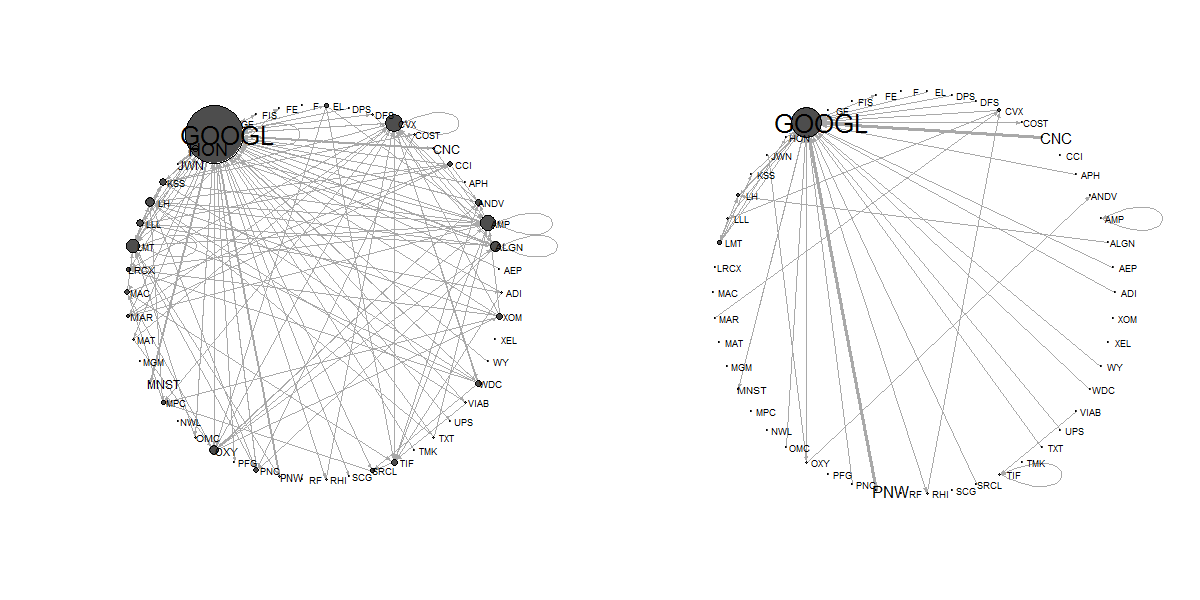}
  \caption{Network visualizations of $VAR(1)$ model parameter estimates for S\&P 500 data using RML (left) and our algorithm (right). Node and label size are proportional to degree centrality.}
  \label{fig2}
\end{figure}

The resulting estimates are visualized in Figure \ref{fig2} as directed graphs comprising nodes representing each vector component and edges indicating the set of nonzero parameters\footnote{That is, $G = (V, E)$ where $V = \{1, \dots, M\}$ and $E = \{(i, j) \in V \times V: A_{ji} \neq 0\}$}. $UoI_{VAR}$ identified just 44 edges that effectively describe a dependence of Google's share price on other companies. Relatively few edges suggest causal networks connecting other share prices. By comparison, RML identifies 146 edges in which the same pattern $UoI_{VAR}$ detected is present but obscured by other edges. One-step forecast RMSE averaged over all companies on the same data for the RML method is 8.3993; for our method, 8.4525 (an increase of $ 0.6\%$ relative to RML). However, the scale of share prices varies widely among the companies; the median per-company forecast RMSE is 4.7807 using RML, and 4.3329 using $UoI_{VAR}$ (a decrease of $9\%$ relative to RML). We conclude that our method finds more interpretable estimates (sparser graph) while maintaining comparable fit.

\section{Discussion}

This paper proposes a novel method ($UoI_{VAR}$) for low-bias and sparse estimation of $VAR$ models, presents simulations that show its advantages, and exemplifies its application in data analysis. The method is flexible, and the hyperparameters allow the analyst to control tolerances for false positives and false negatives without explicitly specifying any \textit{a priori} assumptions about sparsity structure. Promising extensions of this work include: (i) development of analogous methods for multivariate point process models; (ii) theoretical analysis of the algorithm; and (iii) application of the method to scientific datasets.

\section*{Acknowledgments}
The authors thank several anonymous reviewers whose comments improved this work.

\newpage
\bibliographystyle{plainnat}
\bibliography{references}

\end{document}